\documentclass{article}
\usepackage{amsfonts}
\usepackage{amsmath}

\input{tcilatex}
\begin{document}

\title{Effects of Non-Commutativity on Light Hydrogen-Like Atoms and Proton
Radius}
\author{M. Moumni \\
Department of Matter Sciences, University of Biskra; Algeria\\
m.moumni@yahoo.fr \and A. BenSlama \\
Department of Physics, University Constantine1; Algeria\\
a.benslama@yahoo.fr}
\maketitle

\begin{abstract}
We study the corrections induced by the theory of non-commutativity, in both
space-space and space-time versions, on the spectrum of hydrogen-like atoms.
For this, we use the relativistic theory of two-particle systems to take
into account the effects of the reduced mass, and we use perturbation
methods to study the effects of non-commutativity.We apply our study to the
muon hydrogen with the aim to solve the puzzle of proton radius [R. Pohl et
al., Nature 466, 213 (2010) and A. Antognini et al., Science 339, 417
(2013)]. The shifts in the spectrum are found more noticeable in muon $H$ ($%
\mu H$) than in electron $H$ ($eH$) because the corrections depend on the
mass to the third power; This explains the discrepancy between $\mu H$ and $%
eH$ results. In space-space non-commutativity, the parameter required to
resolve the puzzle $\theta _{ss}\approx \left( 0.35GeV\right) ^{-2}$,
exceeds the limit obtained for this parameter from various studies on $eH$
Lamb shift. For space-time non-commutativity, the value $\theta _{st}\approx
\left( 14.3GeV\right) ^{-2}$ has been obtained and it is in agreement with
the limit determined by Lamb shift spectroscopy in $eH$. We have also found
that this value fills the gap between theory and experiment in the case of $%
\mu D$ and improves the agreement between theoretical and experimental
values {}{}in the case of hydrogen-deuterium isotope shift.

\textbf{KeyWords:} Non-Commutativity; H-Like Atoms; Proton Radius

\textbf{PACS:} 02.40.Gh, 67.63.Gh, 31.30.jr
\end{abstract}

\section{Introduction}

Historically, experimental spectroscopy was the perfect test for any theory
having any connection with matter and it played the leading role in
calibrating the values of physical constants. But lately, it has reached
such precision that it accessed the role of indicator of new theories; and
last experience on the Lamb shift in muonic hydrogen \cite{1} is the perfect
example.

The discrepancy between these experimental results $r_{\mu H}=0.84169(66)fm$
and those extracted from electronic hydrogen or elastic electron-proton
scattering and recorded in CODATA $r_{eH}=0.8775(51)fm$ \cite{2} (The values
are at $7\sigma$ variance with respect to each other) has had an impact in
the whole scientific community and this raised many questions about the
cause of such disagreement.

The experimental methods used to obtain the two results are very elaborated.
This is why many studies have investigated how to explain this difference
and to reconcile the two results by trying to rectify the theory. But the
difference between the two experiments remains a puzzle until now and
especially after being reinforced recently with a more accurate value $%
r_{\mu H}=0.84087(39)$ \cite{3}.

Non-perturbative numerical computations of the Dirac equation confirmed the
validity of perturbation methods used to compute the radius \cite{4}-\cite{5}%
. No significant QED correction has been found yet, which would explain the
discrepancy \cite{6}-\cite{7}. Using electron scattering experiments, \cite%
{8} found that data rules out values of the third Zemach moment large enough
to explain the puzzle. Three-body physics does not solve the problem as
demonstrated in \cite{9}. Constraints from low energy data disfavor new
spin-0, spin-1 and spin-2 particles as an explanation \cite{10}. There are
some claims that proton polarizability contribution in the Lamb shift may
explain the discrepancy because it is proportional to the lepton mass to the
fourth power \cite{11}-\cite{12}. These effects could be probed in
scattering experiment planned to run at Paul Scherrer Institute (PSI). For
more information about the different approaches to the problem, see
references \cite{6}-\cite{13}-\cite{14}-\cite{15}.

Because Pohl et al. used an indirect method to calculate $r_{\mu H}$ that
involves comparing the frequency measured experimentally with that given
theoretically according to the radius \cite{1}: 
\begin{subequations}
\begin{align}
\frac{\Delta(2P_{1/2}\rightarrow2S_{1/2})}{meV} & =206.0573(45)-5.2262\frac {%
r_{p}^{2}}{fm^{2}}+0.0347\frac{r_{p}^{3}}{fm^{3}}  \label{1a} \\
\frac{\Delta(2P_{3/2}\rightarrow2S_{1/2})}{meV} & =209.9779(49)-5.2262\frac {%
r_{p}^{2}}{fm^{2}}+0.0347\frac{r_{p}^{3}}{fm^{3}}  \label{1b}
\end{align}
we propose, in this work, to modify the precedent theoretical expressions of
the transition frequency by incorporating the corrections induced by the
non-commutative structure of space-time.

The idea of taking non-commutative space-time coordinates dates from the
thirties of last century. It had as objective to avoid infinities in Coulomb
potentials (gravitation \& electricity) by introducing an lower bound for
the measurement of length. Despite the fact that the concept was suffering
from some problems with unitarity and causality, the theory evolved from the
mathematical point of view, especially after the work of Connes in the
eighties of last century \cite{16}.

In 1999, the work of Seiberg and Witten on string theory \cite{17} has
aroused new interest in the theory. They showed that the dynamics of the
endpoints of an open string on a D-brane in the presence of a magnetic
back-ground field is described by a theory of Yang-Mills on a
non-commutative space-time.

Today, we find non-commutativity in various fields of physics such as solid
state physics, where it was shown that is the framework in which Hall
conductivity is quantized \cite{18} and that it is the proper tool replacing
Bloch's theory whenever translation invariance is broken in aperiodic solids 
\cite{19}. In fluid mechanics, non-commutative fluids are introduced by
studying the quantum Hall effect \cite{20} or bosonization of collective
fermion states \cite{21}. There is also some connection with quantum
statistical physics \cite{22}, and it is also an interpretation of
Ising-type models \cite{23}. One can even find a manifestation of the
non-commutativity in the physiology of the brain, where non-commutative
computation in the vestibulo-ocular reflex was demonstrated in a way that is
unattainable by any commutative system \cite{24}. \cite{25} is an excellent
reference for the different manifestations and applications of
non-commutative field theory.

The theory is a distortion of space-time where the coordinates $x^{%
\mu
}$ become Hermitian operators and thus do not commute: 
\end{subequations}
\begin{equation}
\left[ x_{nc}^{\mu},x_{nc}^{\nu}\right] =i\theta^{\mu\nu};\mu,\nu=0,1,2,3
\label{2}
\end{equation}

The $nc$\ indices denote non-commutative coordinates. $\theta^{%
{\mu}%
\nu}$\ is the parameter of the deformation and it is an anti-symmetric real
matrix. We distinguish two types of non-commutativity; the first one is the
space-space case, where the deformation is introduced between the spatial
coordinates only, and the second is when the spatial coordinates commute
with one another but not with time coordinate and it is noted space-time
case. For a review, one can see reference \cite{26}.

In the literature, there are a lot of studies on hydrogen atom in
non-commutativity. For space-space non-commutativity, we cite \cite{27} \cite%
{28} \cite{29} \cite{30}. For the space-time case, one can see \cite{31} 
\cite{32} \cite{33}.

We have found in \cite{31} and \cite{32}, that the corrections induced by
non-commutativity on the spectrum of the hydrogen atom are proportional to
the lepton mass to the third power (the result is confirmed by \cite{33}),
and this is exactly the shape of the corrections induced by the nuclear size
as demonstrated in \cite{34} \cite{35}. We will apply our result to the
muonic hydrogen, and we will incorporate therein the effects of the finite
mass of the nucleus. We start by computing the corrections to the energies
in both space-space and space-time cases of non-commutativity using
perturbation methods in the Dirac theory of two particles systems. Then we
compare to the difference between theoretical and experimental results
obtained in $\mu H$ experience. This allows us to obtain values of the
non-commutative parameter that resolve the puzzle. Then we will discuss the
possible effects of these corrections on the Lamb shift of muonic deuterium $%
\mu D$ and on the difference between the radii of the proton and deuteron
via the $2S-1S$ transition.

\section{Coulomb Potential in Non-Commutative Space-Time}

We start by rewriting (\ref{2}) for the two versions to consider of the
non-commutativity: 
\begin{subequations}
\begin{align}
\left[ x_{st}^{j},x_{st}^{0}\right] & =i\theta^{j0}  \label{3a} \\
\left[ x_{ss}^{j},x_{ss}^{k}\right] & =i\theta^{jk}  \label{3b}
\end{align}
$st$ subscripts are for space-time case and $ss$ ones are for space-space
case. The $0$ denotes time and Latin indices are used for space coordinates.
To solve these relations, we follow \cite{27} \cite{28} \cite{31} \cite{32}
and choose the Bopp shift formulation of the solutions \cite{36}; we write: 
\end{subequations}
\begin{subequations}
\begin{gather}
x_{st}^{j}=x^{j}-i\theta^{j0}\partial_{0}  \label{4a} \\
x_{ss}^{j}=x^{j}-\frac{i}{2\hbar}\theta^{jk}\partial_{k}  \label{4b}
\end{gather}
The usual coordinates of space $x^{j}$ satisfy the usual canonical
permutation relations and time $x^{0}$ is unchanged in both cases.

We are dealing with the stationary quantum equations, and this allows us to
consider the energy as a constant parameter. The kinetic energy does not
change since it depends on the momentum that remains unchanged, thus we take
the Coulomb potential and construct its non-commutative image. To achieve
this, we have to write the expression of $r_{nc}^{-1}$, where $nc$ denotes
the two cases considered here: 
\end{subequations}
\begin{equation}
\frac{1}{r_{nc}}=\left( \left\Vert \overrightarrow{r}+\overrightarrow {%
\varrho}_{nc}\right\Vert \right) ^{-1}=\left( \tsum x_{nc}^{j}\cdot
x_{nc}^{j}\right) ^{-1/2}  \label{5}
\end{equation}
$\overrightarrow{\varrho}_{nc}$ is the non-commutative correction of
position vector. We make the development in series and because of the
smallness of the non-commutative parameter, as one can see from the bounds
given in the literature \cite{26}, we restrict ourselves to the $1st$ order
in $\theta$ and neglect the higher order terms \cite{27} \cite{31}: 
\begin{subequations}
\begin{align}
r_{st}^{-1} & =\left( 1+i\frac{\partial_{0}\overrightarrow{r}\cdot%
\overrightarrow{\theta}_{st}}{r^{3}}+O(\theta^{2})\right)  \label{6a} \\
r_{ss}^{-1} & =\left( 1+\frac{i}{2}\frac{\left( \overrightarrow{r}\times%
\overrightarrow{\partial}\right) \cdot\overrightarrow{\theta}_{ss}}{r^{3}}%
+O(\theta^{2})\right)  \label{6b}
\end{align}
We have used the vectorial notation: 
\end{subequations}
\begin{subequations}
\begin{align}
\overrightarrow{\theta}_{st} & \equiv\left(
\theta_{st}^{1},\theta_{st}^{2},\theta_{st}^{3}\right)
;\theta_{st}^{j}=\theta^{j0}  \label{7a} \\
\overrightarrow{\theta}_{ss} & \equiv\left(
\theta_{ss}^{1},\theta_{ss}^{2},\theta_{ss}^{3}\right) ;\theta^{jk}=\frac{1}{%
2}\varepsilon^{jkl}\theta_{ss}^{l}  \label{7b} \\
\overrightarrow{\partial} & \equiv\left(
\partial_{1},\partial_{2},\partial_{3}\right)  \label{7c}
\end{align}
Thus, one can write the non-commutative Coulomb potential (up to the $1st$
order $\theta$)\ as follows: 
\end{subequations}
\begin{subequations}
\begin{align}
V_{st}(r) & =-\frac{Ze^{2}}{r}-\frac{Ze^{2}E}{\hbar}\frac{\overrightarrow {r}%
\cdot\overrightarrow{\theta}_{st}}{r^{3}}+O(\theta_{st}^{2})  \label{8a} \\
V_{ss}(r) & =-\frac{Ze^{2}}{r}-\frac{Ze^{2}}{4\hbar}\frac{\overrightarrow {L}%
\cdot\overrightarrow{\theta}_{ss}}{r^{3}}+O(\theta_{ss}^{2})  \label{8b}
\end{align}
where we have used the fact that $i\partial_{0}\psi=H\psi=\left(
E/\hbar\right) \psi$ and $\overrightarrow{r}\times i\hbar\overrightarrow {%
\partial}=\overrightarrow{r}\times\overrightarrow{p}=\overrightarrow{L}$ the
orbital momentum.

An adequate choice of the parameters is $\overrightarrow{\theta}%
_{nc}=\theta^{r0}\overrightarrow{r}/r=\theta_{nc}\overrightarrow{r}/r$. The
writing is similar to that in \cite{37} for space-space non-commutativity
and in \cite{31} and \cite{38} for space-time case. The choice made in this
paper allows us to write the non-commutative Coulomb potential as: 
\end{subequations}
\begin{subequations}
\begin{align}
V_{st}(r) & =-\frac{Ze^{2}}{r}-\frac{Ze^{2}E\theta_{st}}{\hbar}\frac {1}{%
r^{2}}+O(\theta_{st}^{2})  \label{9a} \\
V_{ss}(r) & =-\frac{Ze^{2}}{r}-\frac{Ze^{2}(\overrightarrow{L}\cdot%
\overrightarrow{\theta}_{ss})}{4\hbar}\frac{1}{r^{3}}+O(\theta_{ss}^{2})
\label{9b}
\end{align}

It was indicated in \cite{31} that the effect of space-time noncommutativity
(\ref{9a})\ is similar to the effect of an electric field of a radial dipole
centered on the proton. Similarly, in \cite{30}, the effect of space-space
noncommutativity (\ref{9b}) was presented as equivalent to the effects of
magnetic field or spin.

Now we can compute the corrections induced by this additional term using
perturbative methods in both versions of non-commutativity.

\section{Corrections of the Dirac Energies}

We write the Dirac equation ($\alpha_{i}=\gamma_{0}\gamma_{i}$ and $%
\gamma_{\mu}$\ are the Dirac matrices): 
\end{subequations}
\begin{equation}
i\hbar\partial_{0}=H\psi=\left( \overrightarrow{\alpha}\cdot\overrightarrow {%
p}\right) +m\gamma^{0}+eA_{0}  \label{10}
\end{equation}

After coordinates deformation, we employ the standard Dirac equation but
with the non-commutative Coulomb potential $A_{0}^{(nc)}=-Ze\,/r_{nc}$, so
we get: 
\begin{subequations}
\begin{align}
eA_{0}^{(st)} & =-\frac{Ze^{2}}{r_{st}}=-\frac{Ze^{2}}{r}-\frac {%
Ze^{2}E\theta_{st}}{\hbar}\frac{1}{r^{2}}+O(\theta_{st}^{2})  \label{11a} \\
eA_{0}^{(ss)} & =-\frac{Ze^{2}}{r_{ss}}=-\frac{Ze^{2}}{r}-\frac {Ze^{2}(%
\overrightarrow{L}\cdot\overrightarrow{\theta}_{ss})}{4\hbar}\frac {1}{r^{3}}%
+O(\theta_{ss}^{2})  \label{11b}
\end{align}
As mentioned before, we restrict ourselves to the $1st$ order in $\theta$.
The Hamiltonian can now be expressed as: 
\end{subequations}
\begin{equation}
H=H_{0}+H_{nc}=\left( \overrightarrow{\alpha}\cdot\overrightarrow{p}\right)
+m\gamma^{0}-Ze^{2}/r+\Delta H_{nc}  \label{12}
\end{equation}
$\Delta H_{nc}$ is the non-commutative correction to the usual Dirac
Hamiltonian $H_{0}$: 
\begin{subequations}
\begin{align}
\Delta H_{st} & =-Ze^{2}\left( E/\hbar\right) \theta_{st}r^{-2}  \label{13a}
\\
\Delta H_{ss} & =-Ze^{2}(\overrightarrow{L}\cdot\overrightarrow{\theta}%
_{ss}/4\hbar)r^{-3}  \label{13b}
\end{align}

The smallness of the parameter $\theta$\ allows us to consider
noncommutative corrections with perturbation theory; to the $1st$ order in $%
\theta$, the corrections of the eigenvalues are: 
\end{subequations}
\begin{equation}
\Delta E_{nc}=\left\langle \Delta H_{nc}\right\rangle =\left\langle \Psi(%
\overrightarrow{r})\left\vert \Delta H_{nc}\right\vert \Psi (\overrightarrow{%
r})\right\rangle  \label{14}
\end{equation}
where the $\Psi(\overrightarrow{r})$ are the eingenstates of the Dirac
Hamiltonian for Coulomb potential.

Because the space-time correction is central while space-space one is not
(this is due to the presence of the $\overrightarrow{L}$ operator in $\Delta
H_{ss}$ (\ref{13b})), we will treat the two cases separately.

\subsection{1\textit{st Order Corrections in Space-Time Non-Commutativity}}

To compute the corrections (\ref{14}) with $\Delta H_{nc}$ given by (\ref%
{13a}) and because all the parameters are constants except the coordinate $r$%
, we have:%
\begin{equation}
\Delta E_{st}=\left\langle \Delta H_{st}\right\rangle =-Ze^{2}\left(
E/\hbar\right) \theta_{st}\left\langle \Psi(\overrightarrow{r})\left\vert
r^{-2}\right\vert \Psi(\overrightarrow{r})\right\rangle  \label{15}
\end{equation}
We can use the expression of the $\Psi(\overrightarrow{r})$ from the
literature as done in \cite{30}, or employ the recurrence relations given in 
\cite{39} as we have done in \cite{31}. The obtained mean value for $r^{-2}$
is:%
\begin{equation}
\left\langle \frac{1}{r^{2}}\right\rangle =\frac{2\kappa\left( 2\kappa
\varepsilon-1\right) \left( 1-\varepsilon^{2}\right) ^{3/2}}{Z\alpha \sqrt{%
\kappa^{2}-Z^{2}\alpha^{2}}\left[ 4\left( \kappa^{2}-Z^{2}\alpha ^{2}\right)
-1\right] }\left( \frac{mc}{\hbar}\right) ^{2}  \label{16}
\end{equation}
where $a_{0}=\hbar^{2}/me^{2}$\ is the $1st$ Bohr radius and $\varepsilon
=E/mc^{2}$; $E$\ is the Dirac energy:%
\begin{equation}
E_{n,j}=mc^{2}\left\{ 1+Z^{2}\alpha^{2}\left[ \left( n-j-1/2\right) +\sqrt{%
\left( j+1/2\right) ^{2}-Z^{2}\alpha^{2}}\right] ^{-2}\right\} ^{-1/2}
\label{17}
\end{equation}
$\alpha=e^{2}/\hbar c$ is the fine structure constant and $j=l\pm1/2$\ is
the quantum number associated to the total angular momentum $%
\overrightarrow {j}=\overrightarrow{l}+\overrightarrow{s}$. The number $%
\kappa$\ is giving by the two relations $\kappa=-\left( j+1/2\right) $ if $%
j=\left( l+1/2\right) $ and $\kappa=\left( j+1/2\right) $ if $j=\left(
l-1/2\right) $.

We see that through $\kappa$ in $\left\langle r^{-2}\right\rangle $,
non-commutative corrections removes the degeneracy $j=l+1/2=(l+1)-1/2$ and
acts like the Lamb shift and the energy depends now\ on $(n,j,l)$, unlike
the usual Dirac energies in (\ref{17}). The equivalence between the two
levels of $l$ for the same $j$ in (\ref{17}) is accidental and is due to the
Coulomb potential which is a special case. The additional term in $r^{-2}$
coming from non-commutativity breaks the symmetry and induces the
differences found.

We consider in this study the effects of the finite mass of the nucleus, as
we will discuss electronic and muonic atoms, where the reduced mass varies
by a factor reaching $186$ for hydrogen and $196$ for the deuteron. To
achieve this, we use the solution for two particles Dirac theory \cite{40}
and write the corrected Dirac energies as a shift of the usual ones:

\begin{equation}
E_{n,j}=mc^{2}\left\{ 1+\frac{1}{1+\eta}\left[ \frac{1}{sqr}-1\right] -\frac{%
\eta}{2\left( 1+\eta\right) ^{3}}\left[ \frac{1}{sqr}-1\right] ^{2}\right\}
\label{18}
\end{equation}
Where $\eta=m/M_{N}$\ is the ratio between the orbiting particle mass $m$\
and the nucleus one $M_{N}$ and $sqr$ is the square expression in usual
Dirac energies (\ref{17}):

\begin{equation}
sqr=\sqrt{1+Z^{2}\alpha^{2}\left[ \left( n-j-1/2\right) +\sqrt{\left(
j+1/2\right) ^{2}-Z^{2}\alpha^{2}}\right] ^{-2}}  \label{19}
\end{equation}
One retrieves (\ref{17}) by putting $\eta=0$\ in (\ref{18}).

We use the shift (\ref{18}) in the expression of $\left\langle
r^{-2}\right\rangle $ when computing the non-commutative corrections to the
energies and thereby we generalize the recurrence relations of \cite{39}\
for the case of a relativistic two-particle system; the result writes:%
\begin{equation}
\Delta E_{n,j^{\pm }}^{(st)}=f^{(st)}(n,j^{\pm },Z\alpha ,\eta )\left(
m^{3}Ze^{2}c^{4}/\hbar ^{3}\right) \theta _{st}  \label{20}
\end{equation}%
Here $f^{(st)}(n,j^{\pm },Z\alpha ,\eta )$ is a dimensionless coefficient
dependent on the parameters within the parentheses ($j^{\pm }$ means $j=l\pm
1/2$). To give an overview of this dependence, we develop its expression
according to $Z\alpha $, and we limit ourselves to the $4th$ order: 
\begin{subequations}
\begin{gather}
f_{j^{+}}^{(st)}=\tfrac{-Z^{2}\alpha ^{2}}{\left( 1+\eta \right) ^{3/2}jn^{3}%
}\left[ 1+\left( \tfrac{6j^{2}+6j+1}{j\left( j+1\right) \left( 2j+1\right)
^{2}}+\tfrac{3}{\left( 2j+1\right) n}-\tfrac{9\eta ^{2}j+26\eta j+9\eta
^{2}+24\eta +20j+18}{8\left( 1+\eta \right) ^{2}\left( j+1\right) n^{2}}%
\right) Z^{2}\alpha ^{2}\right]  \label{21a} \\
f_{j^{-}}^{(st)}=\tfrac{-Z^{2}\alpha ^{2}}{\left( 1+\eta \right)
^{3/2}(j+1)n^{3}\hbar ^{3}}\left[ 1+\left( \tfrac{6j^{2}+6j+1}{j\left(
j+1\right) \left( 2j+1\right) ^{2}}+\tfrac{3}{\left( 2j+1\right) n}-\tfrac{%
9\eta ^{2}j+26\eta j+2\eta +20j+2}{8\left( 1+\eta \right) ^{2}jn^{2}}\right)
Z^{2}\alpha ^{2}\right]  \label{21b}
\end{gather}%
When putting $\eta =0$\ and $Z=1$ in (\ref{20}), (\ref{21a}) and (\ref{21b}%
), we recover the results of \cite{31}.

This concludes our study of the space-time case.

\subsection{\textit{1st Order Corrections in Space-Space Non-Commutativity}}

For the space-space case and as mentioned before, the corrections are no
longer central (\ref{13b}) and we must use the full expression of the
spinors $\Psi(\overrightarrow{r})$ to perform the computations (we cite for
example \cite{41} and \cite{42}). The method is very time consuming but the
work has already been done in \cite{28} and \cite{30}. For the level in
which we are interested in this study ($n=2$), the corrections were found
proportional to $m^{3}c^{4}Z^{4}\alpha^{4}/\hbar^{2}$; we write their values
from \cite{28} ($Z=1$): 
\end{subequations}
\begin{subequations}
\begin{align}
\Delta E_{2S_{1/2}}^{(ss)} & =0  \label{22a} \\
\Delta E_{2P_{1/2}}^{(ss)} & =\pm6.75\times10^{19}\theta_{ss}eV/m^{2}
\label{22b} \\
\Delta E_{2P_{3/2}}^{(ss)} & =\pm\left( m_{l}+\tfrac{1}{2}\right)
\,6.75\times10^{19}\theta_{ss}eV/m^{2}  \label{22c}
\end{align}
where $\theta_{ss}$ is in $m^{2}$ and $m_{l}=0,1$.

The sign $\pm$\ comes from the fact that $\Delta H_{ss}$\ removes the
degeneracy according to the azimuthal quantum number $m_{l}$\ via the
operator $\overrightarrow{L}$, and the $1/2$ arises because the $\Psi(%
\overrightarrow {r})$ for Dirac equation, are eingenstates of the operator $%
\overrightarrow {J}=\overrightarrow{L}+\frac{1}{2}diag(\overrightarrow{\sigma%
},\overrightarrow {\sigma})$ instead of $\overrightarrow{L}$.

We propose here to give another method to estimate the corrections in the
space-space case. We start as done in \cite{28} and \cite{30}, by writing: 
\end{subequations}
\begin{equation}
\left\langle \Delta H_{ss}\right\rangle =-\frac{Ze^{2}}{4\hbar}\left\langle
\Psi(r)\left\vert r^{-3}\right\vert \Psi(r)\right\rangle \left\langle
\Psi(\vartheta,\varphi)\left\vert \overrightarrow{L}\cdot\overrightarrow {%
\theta}_{ss}\right\vert \Psi(\vartheta,\varphi)\right\rangle  \label{23}
\end{equation}

The term in angular coordinates is a matrix, when diagonalized gives a
contribution proportional to $\theta_{ss}=\sqrt{\theta_{j}\theta^{j}}$,\ and
the proportion is a simple fraction noted $a$, so we approximate this term
to $a\theta_{ss}$. We justify this by noting that when neglecting the
fine-structure constant $\alpha$ compared to unity, the $\Psi(%
\overrightarrow {r})$ in Dirac equation becomes precisely the normalized Schr%
\"{o}dinger eigenfunction provided we express the parameter $\kappa$ in
terms of $l$ in the relativistic functions \cite{41} and this gives the
parameter $\pm\left( m_{l}+\frac{1}{2}\right) $ in $\Delta E^{(ss)}$.

For the radial term, and as done in space-time case, we use recurrence
relation from \cite{39}:%
\begin{equation}
\left\langle \frac{1}{r^{3}}\right\rangle =\frac{2\left[ 3\kappa
^{2}\varepsilon^{2}-3\kappa\varepsilon-\left( \kappa^{2}-Z^{2}\alpha
^{2}\right) +1\right] \left( 1-\varepsilon^{2}\right) ^{3/2}}{\sqrt {%
\kappa^{2}-Z^{2}\alpha^{2}}\left[ \left( \kappa^{2}-Z^{2}\alpha^{2}\right) -1%
\right] \left[ 4\left( \kappa^{2}-Z^{2}\alpha^{2}\right) -1\right] }\left( 
\frac{mc}{\hbar}\right) ^{3}  \label{24}
\end{equation}
with the same notations as in (\ref{16}).

In this case too, we consider the effects of the finite mass of the nucleon.
When using the shift (\ref{18}) in the expression of $\left\langle
r^{-3}\right\rangle $, the non-commutative corrections to the energies are:%
\begin{equation}
\Delta E_{n,j^{\pm}}^{(ss)}=f^{(ss)}(n,j^{\pm},Z\alpha,\eta)\left(
m^{3}Ze^{2}c^{3}/\hbar^{3}\right) a\theta_{st}  \label{25}
\end{equation}
with the same notations as in space-time case. The approximate solution of
the coefficient $f$ is: 
\begin{subequations}
\begin{gather}
f_{j^{+}}^{(ss)}=\tfrac{\left( 1+\eta\right) ^{-3/2}Z^{3}\alpha^{3}}{j\left(
2j-1\right) \left( 2j+1\right) n^{3}}\left[ 1+\left( 
\begin{array}{c}
\tfrac{48j^{2}+36j^{2}-8j-3}{j\left( 2j-1\right) \left( 2j+1\right)
^{2}\left( 2j+3\right) }+\tfrac{3}{\left( 2j+1\right) n} \\ 
-3\tfrac{6\eta^{2}j+20\eta j+9\eta^{2}+22\eta+16j+16}{8\left( 1+\eta\right)
^{2}\left( 2j+3\right) n^{2}}%
\end{array}
\right) Z^{2}\alpha^{2}\right]  \label{26a} \\
f_{j^{-}}^{(ss)}=\tfrac{\left( 1+\eta\right) ^{-3/2}Z^{3}\alpha^{3}}{%
(j+1)(2j+1)(2j+3)n^{3}}\left[ 1+\left( 
\begin{array}{c}
\tfrac{48j^{2}+108j^{2}+64j+7}{\left( j+1\right) \left( 2j-1\right) \left(
2j+1\right) ^{2}\left( 2j+3\right) }+\tfrac{3}{\left( 2j+1\right) n} \\ 
-3\tfrac{6\eta^{2}j+20\eta j-3\eta^{2}-2\eta+16j}{8\left( 1+\eta\right)
^{2}\left( 2j-1\right) n^{2}}%
\end{array}
\right) Z^{2}\alpha^{2}\right]  \label{26b}
\end{gather}

To justify our approach, we apply it to the level $n=2$ of the hydrogen and
we find ($Z=1$): 
\end{subequations}
\begin{subequations}
\begin{align}
\Delta E_{2S_{1/2}}^{(ss)} & =\pm9.11531\times10^{-4}\left(
m^{3}e^{2}c^{3}/\hbar^{3}\right) a\theta_{ss}  \label{27a} \\
\Delta E_{2P_{1/2}}^{(ss)} & =\pm5.05790\times10^{-9}\left(
m^{3}e^{2}c^{3}/\hbar^{3}\right) a\theta_{ss}  \label{27b} \\
\Delta E_{2P_{3/2}}^{(ss)} & =\pm4.04467\times10^{-9}\left(
m^{3}e^{2}c^{3}/\hbar^{3}\right) a\theta_{ss}  \label{27c}
\end{align}

For the $2S_{1/2}$\ state, we put $a=0$\ because the $\overrightarrow{L}$
operator vanishes in this state and we retrieve the result of \cite{28}
here. For the $P$-states, comparing our results to those coming from \cite%
{28} gives: 
\end{subequations}
\begin{equation}
a\left( 2P_{1/2}\right) =0.53\,\&\,a\left( 2P_{3/2}\right) =0.33-1.00
\label{28}
\end{equation}

These values show that our method gives results very close to the exact
values from \cite{28} and \cite{30} but it is much less tedious. It also has
the advantage of providing a general formulation of the results, which is
not the case of the pre-cited studies.

\section{Non-Commutativity in H-Like Atoms}

Now we apply the expressions of non-commutative corrections found in the
previous section on hydrogen and deuterium whether electronic or muonic,
with the aim to find an explanation for the different results of
aforementioned experiments. We will start with the result that has generated
the more debates and that relates to the muonic hydrogen. Then we will
discuss some aspects concerning the radii of proton and deuteron and atomic
spectroscopy through muonic deuterium, electronic hydrogen and electronic
deuterium.

\subsection{\textit{Muon Hydrogen and Non-Commutativity}}

In this section, we will apply our study of both space-time and space-space
cases of non-commutativity to the muon hydrogen and especially to the
transitions used to compute the charge radius of the proton (\ref{1a}) and (%
\ref{1b}). Using the value of the charge radius given by CODATA 2010 $%
0.8775(51)fm$ \cite{2}, the results obtained for the precedent transitions
differ from those found in experiments on muonic hydrogen \cite{1}, by an
amount equal to $0.32meV$. Although the difference is very small, it is
significant given the precision of the experiments used.

We compute the non-commutative corrections to these transitions for the
space-time case using (\ref{16}) (\ref{17}) (\ref{18}) and (\ref{19}); we
find : 
\begin{subequations}
\begin{align}
\Delta E_{2P_{1/2}}^{(st)}-\Delta E_{2S_{1/2}}^{(st)} & =6.51015\times
10^{16}\left( \theta_{st}eV^{2}\right) eV  \label{29a} \\
\Delta E_{2P_{3/2}}^{(st)}-\Delta E_{2S_{1/2}}^{(st)} & =6.51044\times
10^{16}\left( \theta_{st}eV^{2}\right) eV  \label{29b}
\end{align}
For the space-space type, we use (\ref{19}) (\ref{23}) (\ref{24}) and (\ref%
{25}): 
\end{subequations}
\begin{subequations}
\begin{align}
\Delta E_{2P_{1/2}}^{(ss)}-\Delta E_{2S_{1/2}}^{(ss)} & =3.93686\times
10^{13}\left( \theta_{ss}eV^{2}\right) eV  \label{30a} \\
\Delta E_{2P_{3/2}}^{(ss)}-\Delta E_{2S_{1/2}}^{(ss)} & =2.96894\times
10^{13}\left( \theta_{ss}eV^{2}\right) eV  \label{30b}
\end{align}
We have chosen $a\left( 2S_{1/2}\right) =0,a\left( 2P_{1/2}\right)
=0.5\,\&\,a\left( 2P_{3/2}\right) =1$ to get back the results of \cite{28}.

Comparing these results to the deviation $0.32\times10^{-3}eV$, we compute
the values of the parameter of non-commutativity that is required to fill
the gap: 
\end{subequations}
\begin{subequations}
\begin{align}
\Delta E_{2P_{1/2}}^{(st)}-\Delta E_{2S_{1/2}}^{(st)} &
=0.32meV\Longrightarrow\theta_{st}=\left( 14.286GeV\right) ^{-2}  \label{31a}
\\
\Delta E_{2P_{3/2}}^{(st)}-\Delta E_{2S_{1/2}}^{(st)} &
=0.32meV\Longrightarrow\theta_{st}=\left( 14.286GeV\right) ^{-2} \\
\Delta E_{2P_{1/2}}^{(ss)}-\Delta E_{2S_{1/2}}^{(ss)} &
=0.32meV\Longrightarrow\theta_{ss}=\left( 0.305GeV\right) ^{-2} \\
\Delta E_{2P_{3/2}}^{(ss)}-\Delta E_{2S_{1/2}}^{(ss)} &
=0.32meV\Longrightarrow\theta_{ss}=\left( 0.351GeV\right) ^{-2}
\end{align}

The value obtained in the case of space-space non-commutativity $\theta
_{ss}\approx\left( 0.3GeV\right) ^{-2}$ exceeds the limit obtained for this
parameter from studies on $eH$ Lamb shift $\theta_{ss}\leq\left(
0.6GeV\right) ^{-2}$ \cite{33}. If we use this limit to compute the radius,
we find $0.86409fm$ which is outside the experimental limits for $r_{\mu H}$ 
\cite{1} and \cite{3}; So this effect is ruled out.

For the case of the space-time non-commutativity, the value $\theta
_{st}\approx \left( 14.3GeV\right) ^{-2}$ is obtained and it is in agreement
with the limit $\theta _{st}\leq \left( 6GeV\right) ^{-2}$ determined by
Lamb shift spectroscopy in $eH$ \cite{33}. It should also be noted that the
corrections are the same for both levels $2P_{1/2}$ and $2P_{3/2}$; this is
in agreement with the two relations (\ref{1a}) and (\ref{1b}), where the
terms in $r$ are equal for both transitions.

This ends our quantitative analysis of non-commutative effects on the
spectrum of muonic hydrogen and on the charge radius of the proton.

\subsection{\textit{Lamb Shift in Muon Deuterium}}

Now, we study the consequences of the noncommutative correction found for $%
\mu H$ on the muon deuterium $\mu D$. The $2P-2S$ Lamb shift has been
extensively studied to see if the results obtained with $\mu H$ are
confirmed by this new muon system. The theoretical value of this transition
depends on the deuteron radius and is given by the formula \cite{43}: 
\end{subequations}
\begin{subequations}
\begin{equation}
\frac{\Delta(2P_{1/2}\rightarrow2S_{1/2})}{meV}=230.2972(400)-6.10940\frac {%
r^{2}}{fm^{2}}+0.0448\frac{r^{3}}{fm^{3}}  \label{32}
\end{equation}

The Mainz collaboration has studied the Lamb-shift of $\mu D$ and found that
the experimental value differs from the theoretical one by $0.383meV$ \cite%
{44} (and references therein). Although the difference is only $3\sigma$,
there is scope for study in this phenomenon. Several studies have looked at
this problem and especially on the possible correction to the two-photon
exchange (proportional to $r^{3}$) to find a solution, but without
conclusive results until now \cite{45} \cite{46} (For a review one can see 
\cite{47}).

We propose to solve this problem by using the same procedure used to $\mu H$%
. To do this, we calculate the non-commutative correction to the transition
for space-time case. From (\ref{20}), (\ref{21a}) and (\ref{21b}), the
non-commutative correction to the Lamb shift in hydrogen-like atoms is given
with the general expression: 
\end{subequations}
\begin{gather}
\Delta E_{n,j}^{(st)}(Lamb\ shift)=\Delta E_{n,j=(l+1)-1/2}^{\left(
st\right) }-\Delta E_{n,j=l+1/2}^{\left( st\right) }  \notag \\
=\tfrac{1}{\left( 1+\eta \right) ^{3/2}}\tfrac{m^{3}c^{4}e^{2}Z^{3}\alpha
^{2}\theta _{st}}{j(j+1)n^{3}\hbar ^{3}}\left[ 1+\left( \tfrac{6j^{2}+6j+1}{%
j\left( j+1\right) \left( 2j+1\right) ^{2}}+\tfrac{3}{\left( 2j+1\right) n}-%
\tfrac{9\eta ^{2}+22\eta +16}{8\left( 1+\eta \right) ^{2}n^{2}}\right)
Z^{2}\alpha ^{2}\right]  \label{33}
\end{gather}

For the $2P_{1/2}\rightarrow2S_{1/2}$ case in $\mu D$, we take $\eta=m_{\mu
}/M_{D}$, $n=2$ and $j=1/2$; we find:%
\begin{equation}
\Delta E_{2,1/2}^{(st)}(LS)=7.03726\times10^{16}\left(
\theta_{st}eV^{2}\right) eV  \label{34}
\end{equation}

The same value is obtained when using the exact expressions (\ref{20}).
Putting the value $\theta_{st}\approx\left( 14.3eV\right) ^{-2}$ (\ref{31a}%
)\ obtained from $\mu H$\ in (\ref{34}), we get $\Delta E^{(st)}=0.348meV$;
this value is approximately equal to the discrepancy between theoretical and
experimental results $0.383meV$. We see that the same parameter corrects the
two observed discrepancies in both $\mu H$ and $\mu D$.

Be noted that the value $0.348meV$\ in $\mu D$ is very close the value $%
0.32meV$ that counts for $\mu H$ and this is easily explained by the fact
that the difference between the two systems comes from the factor $%
m^{3}\left( 1+\eta\right) ^{-3/2}$ in (\ref{33}) and the ratio of\ the
numerical quantities is $0.93\approx1$.

\subsection{\textit{2S-1S Transition in Hydrogen and Deuterium}}

It should be noted that the difference between the radii of proton and
deuteron is a very well-defined parameter using the $2S-1S$ transition which
is one of the most accurate measured quantities \cite{48}. This transition
is used because the effects due to nuclear size ($ns$) are, to the first
order, nonzero only for these states \cite{34}:%
\begin{equation}
\Delta E_{ns}=\frac{2}{3}\frac{\mu^{3}c^{4}Z^{4}\alpha^{4}}{\hbar^{2}n^{3}}%
\left\langle r^{2}\right\rangle \delta_{l0}  \label{35}
\end{equation}
We see that knowing the reduced masses $\mu$ allows us to accurately
evaluate radii of nuclei from high precision spectroscopy.

The most recent measurement of the hydrogen-deuterium isotope shift is \cite%
{49}:%
\begin{equation}
\Delta f_{ex}(H-D)=f_{H}^{1S-2S}-f_{D}^{1S-2S}=670994334606(15)Hz  \label{36}
\end{equation}

and the most recent theoretical evaluation is \cite{50}:%
\begin{equation}
\Delta f_{th}(H-D)=f_{H}^{1S-2S}-f_{D}^{1S-2S}=670994346(23)kHz  \label{37}
\end{equation}
Both values agree well within the limits given.

We will evaluate the contribution coming from space-time non-commutativity
to this shift. We use the relations from (\ref{15}) to (\ref{20}), to
compute the non-commutative corrections to the transition in both hydrogen
and deuterium: 
\begin{subequations}
\begin{align}
\Delta E_{2S}^{(st)}(H)-\Delta E_{1S}^{(st)}(H) &
=hf_{H}^{1S-2S}=9.06690\times10^{10}\left( \theta_{st}eV^{2}\right) eV
\label{38a} \\
\Delta E_{2S}^{(st)}(D)-\Delta E_{1S}^{(st)}(D) &
=hf_{D}^{1S-2S}=9.07060\times10^{10}\left( \theta_{st}eV^{2}\right) eV
\label{38b}
\end{align}

Inserting the value of the parameter $\theta _{st}$ found for $\mu H$ (\ref%
{31a}) in these expressions, we find the non-commutative correction to the
hydrogen-deuterium isotope shift: 
\end{subequations}
\begin{equation}
\Delta f_{th}^{(nc)}(H-D)=f_{H}^{1S-2S}-f_{D}^{1S-2S}=275.488Hz  \label{39}
\end{equation}%
This contribution don't fill the gap of $12kHz$\ between $\Delta f_{th}$\
and $\Delta f_{ex}$, but it improves slightly the agreement between the two
values. This confirms the fact that there are no doubts about the results of
experiments on $eH$ spectroscopy because they are so accurate and one has to
look the side of muonic systems.

\section{Conclusion}

In this work, we studied the corrections induced by a non-commutative
structure of space-time, in its two versions space-space and space-time, on
the spectrum of hydrogen-like atoms. We have applied our study to the muonic
hydrogen and this with the aim to solve the puzzle of proton radius, because
we think that the experiments used to study this phenomenon are so developed
that we can not doubt of their results, and therefore one must look on the
side of theory of atomic spectroscopy used to compute the radius.

In this study, we considered the effects of the mass of the nucleus of the
non-commutative corrections and thus we have improved previous works in this
area. This allowed us to consider the difference caused by changing the
nucleus (from proton to deuteron) in addition to that which occurs when
changing the orbiting particle (from electron to muon).

It should be noted that the effects of the nucleus shape on the energy
levels of the atom are proportional to the third power of the mass of the
orbiting particle; this is easily understood by the fact that the Bohr
radius ($a_{0}=\hbar^{2}/me^{2}$) is inversely proportional to the mass and
thus the particle is that much nearer the nucleus, that its mass is greater;
and this makes it more sensitive to these effects.

We have demonstrated that the effects of non-commutativity are also
proportional to the third power of the mass of the particle because it
distorts the Coulomb potential and adds a term proportional to $r^{-2}$ in
space-time case and to $r^{-3}$ in space-space case. It is for this reason
also that the effects decreases with increasing values of quantum numbers as
can be seen in the different relations of the spectrum corrections (because
the term $r^{-n}$ with $n>1$ is very steep for small values of $r$). This is
why we use this theory to explain the puzzle because its effects are
different depending on whether it is applied to muonic hydrogen or
electronic hydrogen. The shifts in the spectrum are more noticeable in muon $%
H$ than in ordinary ones, and this explains the fact that experiments on $%
\mu H$ spectroscopy give results that are different from those obtained with 
$eH$.

In the case of space-space non-commutativity, the parameter required to
resolve the puzzle is $\theta_{ss}\approx\left( 0.35GeV\right) ^{-2}$. This
value exceeds the limit obtained for this parameter from studies on $eH$
Lamb shift $\theta_{ss}\leq\left( 0.6GeV\right) ^{-2}$ \cite{33}. If we use
this limit to compute the radius, we find $0.86409fm$ which is outside the
experimental limits for $r_{\mu H}$ \cite{1} and \cite{3}. Another problem
arises with in this case; it is the $\pm$ sign in the corrections (due to
the presence of the azimuthal quantum number in their expressions). This
sign means that the corrected value of the radius ranges from $0.84169fm $
to $0.91192fm$ in violation of $\mu H$ experiences.

In the case of the space-time non-commutativity, the value $\theta
_{st}\approx \left( 14.3GeV\right) ^{-2}$ has been obtained and it is in
agreement with the limit determined by Lamb shift spectroscopy in $eH$ $%
\theta _{st}\leq \left( 6GeV\right) ^{-2}$ \cite{33}. It was also found in
this case, that the corrections are the same for both levels $2P_{1/2}$ and $%
2P_{3/2}$ although we found that corrections remove the degeneracy of the
Dirac energies\ with respect to the total angular momentum quantum number $%
j=l+1/2=(l+1)-1/2$ (non-commutativity acts like the Lamb Shift here). This
is in agreement with the two relations (\ref{27a}) and (\ref{27b}), where
the terms in $r$ are equal for both transitions. This is not true for the
space-space case because the corrections of the two levels differ from one
another (\ref{29a}) (\ref{29b}). We say that this is a consequence of the
fact that the correction term to the Coulomb potential is proportional to $%
r^{-2}$ in the space-time case, and so as we have previously mentioned in 
\cite{31}, we assimilate it to the field of a central dipole. In other
words, the action of space-time non-commutativity is equivalent to consider
the extended charged nature of the proton in the nucleus, which is the
principal characteristic studied in $\mu H$ experiments.

When applying the result obtained from the study of $\mu H$ in $\mu D$, we
found a correction of $0.348meV$ which is almost exactly equal to the
difference between theory and experiment for this system. The very close
values of the corrections in these two systems $\mu H$ and $\mu D$ are
easily explained by the fact that the ratio between the reduced masses, of
the two is $0.95\approx1$ (The disagreements between theory and experiment
in both systems are almost equal).

Using the same result coming from $\mu H$ for the $eH$-$eD$ isotope shift,
the correction found improves the agreement between theoretical and
experimental results (which was already excellent). The same reasoning as
above is used, and we say that the corrections in electronic systems either $%
eH$ or $eD$ are infinitely small compared to those of muonic systems; the
ratio between the reduced masses of the two is $\approx10^{-7}$.

Eyes are now turned to the results of experimental on $\mu p$ scattering and 
$\mu He$ spectroscopy to see whether the phenomenon is spectroscopic or is
it due to the nature of the particles. On the side of electronic systems,
there is practically no doubt on their veracity; the radius of the proton
was even calculated in a model independent way from $ep$ scattering \cite{51}
\cite{52} and the results confirm CODATA value.

It was reported in our work \cite{31}, that the limit on the parameter $%
\theta _{st}\approx \left( 1TeV\right) ^{-2}$, and so our result here is
greater than the latter; however it was pointed out to us that such a limit (%
$\theta _{st}\approx \left( 1TeV\right) ^{-2}$) should not only be estimated
according to experimental precision calculations but rather on the
disagreement between experiment and theory, which requires a correction of
the limit of \cite{31} and this is what is done in this work for the muon
hydrogen.

We can also evoke that the proton raises other questions about its
properties in addition to the one discussed in this article and we can
mention as an example the origin of its spin or what is called "spin crisis"
in \cite{53} or "proton spin puzzle "in \cite{54}.

\end{document}